\documentclass[]{spie}  %>>> use for US letter paper
 \usepackage{booktabs}
\usepackage{amsmath,amsfonts,amssymb}
\usepackage{graphicx}
\usepackage[colorlinks=true, allcolors=blue]{hyperref}
\usepackage{comment}

\title{The XGIS instrument on-board THESEUS:\\ the detection plane and on-board electronics}

\author[a,b]{F.~Fuschino}
\author[a,b]{R.~Campana}
\author[a]{C.~Labanti}
\author[a]{L.~Amati}
\author[a,b]{E.~Virgilli}
\author[a]{L.~Terenzi}
\author[c]{P.~Bellutti}
\author[d]{G.~Bertuccio}
\author[c]{G.~Borghi}
\author[g]{L.~C.~Bune~Jensen}
\author[c]{F.~Ficorella}
\author[d]{M.~Gandola}
\author[e]{A.~Gemelli}
\author[e]{M.~Grassi}
\author[f]{P.~Hedderman}
\author[g]{I.~Kuvvetli}
\author[h]{G.~La~Rosa}
\author[i]{P.~Lorenzi}
\author[e]{P.~Malcovati}
\author[d]{F.~Mele}
\author[l]{P.~Orleanski}
\author[g]{M.~Pedersen}
\author[c]{A.~Picciotto}
\author[m]{A.~Rachevski}
\author[m,n]{I.~Rashevskaya}
\author[f]{A.~Santangelo}
\author[i]{P.~Sarra}
\author[h]{G.~Sottile}
\author[g]{D.~Tcherniak}
\author[f]{C.~Tenzer}
\author[o,n]{A.~Vacchi}
\author[l]{M.~Winkler}
\author[n]{G.~Zampa}
\author[n]{N.~Zampa}
\author[c]{N.~Zorzi}

\affil[a]{INAF/OAS, Via Gobetti 101, I-40129, Bologna, Italy}
\affil[b]{INFN-Sezione di Bologna, Viale Berti Pichat 6/2, I-40127 Bologna, Italy}
\affil[c]{Fondazione Bruno Kessler – FBK, Via Sommarive 18, I-38123 Trento, Italy}
\affil[d]{Department of Electronics, Information and Bioengineering (DEIB) of Politecnico di Milano, Como Campus, Via Anzani 42, 22100 Como, Italy}
\affil[e]{Department of Electrical, Computer, and Biomedical Engineering, University of Pavia, Via Ferrata 5, 27100, Pavia, Italy}
\affil[f]{Institut f{\"u}r Astronomie und Astrophysik, Abteilung Hochenergieastrophysik, Kepler Center for Astro and Particle Physics, Eberhard Karls Universit{\"a}t T{\"u}bingen, Sand 1, 72076 T{\"u}bingen, Germany}
\affil[g]{National Space Institute, Technical University of Denmark, Elektrovej building 327, Denmark}
\affil[h]{INAF, Istituto di Astrofisica Spaziale e Fisica cosmica di Palermo, via U. La Malfa 153, I-90146 Palermo, Italy}
\affil[i]{OHB-Italia, Via Gallarate, 150, I-20151 Milano, Italy}
\affil[l]{Space Research Centre, Polish Academy of Sciences, Bartycka 18A, 00-716 Warszawa, Poland}
\affil[m]{TIFPA-INFN, Via Sommarive 14, I-38123 Trento, Italy}
\affil[n]{INFN Italian National Institute for Nuclear Physics 
c/o Area di Ricerca, Padriciano 99, I-34127, Trieste, Italy}
\affil[o]{Department of Mathematics, Computer Science and Physics
University of Udine, Via delle Scienze 206, I-33100, Udine, Italy}

\authorinfo{Send correspondence to:\\
Fabio Fuschino, E-mail: \url{fabio.fuschino@inaf.it}\\
}

\begin{document} 
\maketitle

\begin{abstract}

The X and Gamma Imaging Spectrometer instrument on-board the THESEUS mission (selected by ESA in the framework of the Cosmic Vision M5 launch opportunity, currently in phase A) is based on a detection plane composed of several thousands of single active elements. Each element comprises a 4.5$\times$4.5$\times$30~mm$^3$ CsI(Tl) scintillator bar, optically coupled at both ends to Silicon Drift Detectors (SDDs). The SDDs acts both as photodetectors for the scintillation light and as direct X-ray sensors. In this paper the design of the XGIS detection plane is reviewed, outlining the strategic choices in terms of modularity and redundancy of the system. Results on detector-electronics prototypes are also described. Moreover, the design and development of the low-noise front-end electronics is presented, emphasizing the innovative architectural design based on custom-designed Application-Specific Integrated Circuits (ASICs).

\end{abstract}

\keywords{THESEUS mission, Chipset, ORION, ASIC, Silicon Drift Detector, Scintillator crystals}

\section{INTRODUCTION}
\label{intruduction}

%cita Lorenzo

% Architettura ASIC
% Architettura Back-end
% Report di avanzamento
% immagine camera

The \emph{Transient High-Energy Sky and Early Universe Surveyor} (THESEUS\cite{amati18, amati2020spieB}) is a space mission concept aiming at fully exploiting Gamma-Ray Bursts (GRBs) for investigating the early Universe and at providing a substantial advancement of multi-messenger and time-domain astrophysics. THESEUS is currently in Phase A study by the European Space Agency (ESA), as a candidate mission for the M5 slot for a possible launch in 2032. 
The Phase A study will be completed
in Spring 2021 with the final downselection to 
one candidate in Summer 2021. The THESEUS mission is 
planned to be launched in a low Earth orbit (LEO) with 
low inclination ($<$6$^\circ$). Its planned nominal lifetime is 4 years,
which ensures a sufficient number of high-redshift GRBs 
(according to the yearly rate of those type of events, which are among the main target of the mission) to be observed and studied.

\begin{figure} [ht]
\centering
\begin{tabular}{c}
\includegraphics[height=10cm]{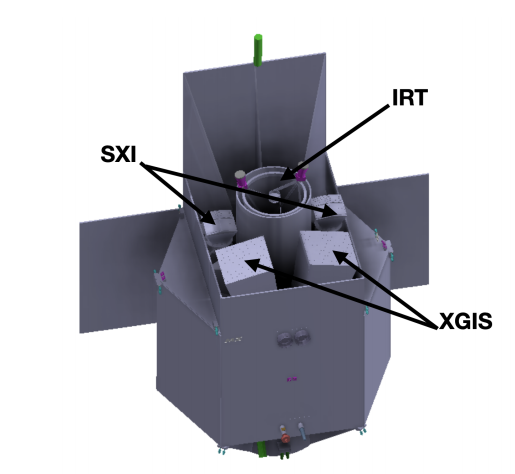}
\end{tabular}
\caption[example] 
{\label{fig:theseus-mission} 
An engineering drawings view of the THESEUS mission with the three on-board instruments.}
\end{figure}

The THESEUS payload combines two high energy instruments
with large a Field of View (FoV) working in synergy for transient events detection. The \emph{X-Gamma Imaging Spectrometer}\cite{amati19} (XGIS) is a wide field deep sky monitor covering the 2 keV--10 MeV  energy pass-band. The XGIS consists of two units, with imaging capability in the 2--150~keV band through the employ of a coded aperture mask. In this energy range the FoV is 77$^\circ$ $\times$ 117$^\circ$ while above $\sim$150~keV  the instrument acts as a full sky spectrometer. The two XGIS units are pointed at $\pm$20$^\circ$ offset with respect to the payload axis in a way that their FoV partially overlap. 
Extensive study of the expected sources of background on the XGIS has been done and in particular of its effects 
on the instrument sensitivity\cite{campana2020spie}~.
The THESEUS \emph{Soft X-ray Imager}\cite{obrien19, obrien18a} (SXI) comprises 2 detector units (DU). Its 0.3--5~keV energy pass band partially overlaps the XGIS energy range. Each DU is a wide field Lobster Eye telescope using the optical principle first described by Angel (1979)\cite{Angel79}~. The optics aperture is formed by an array of 8$\times$8 square pore Micro Channel Plates (MCPs). The MCPs are 40$\times$40~mm$^2$ and are mounted on a spherical frame with radius of curvature 600~mm (2 times the focal length of 300~mm). The use of the Lobster Eye concept allows an unprecedented combination of large FoV (0.5~sr), source location accuracy ($<$~1--2~arcmin). The SXI and XGIS triggers on transient events are exploited by the third instrument of THESEUS, the near \emph{Infra-Red Telescope}\cite{gotz17, gotz18} (IRT) which is
composed by a primary mirror of 0.7~m of diameter and a secondary mirror of 0.23~m in a Cassegrain configuration. In imaging mode the IRT has a FoV of   
$15'\times15'$. Transient events that triggers in the 
FoV of the SXI and/or XGIS will be observed by IRT that 
will provide immediate identification systematically 
within a few minutes with sub-arcsecond position accuracy 
and spectroscopic redshift determination.
An overview of the THESEUS mission with its 
payload is given in Figure~\ref{fig:theseus-mission}. 
The three instruments are coaxial and in particular, the SXI FoV is fully included in that of the XGIS.

In the following Sections a detailed description of the XGIS will be 
given, focusing on the design of the 
detection plane and on the related readout electronics.
For each XGIS camera, a customized architecture 
for the electronics is required, given  the 
complexity and peculiarity of the detection plane, 
made with 12800 silicon-based detectors coupled in pairs to 6400 scintillator bars.  
The design of an Application Specific Integrated Circuit (ASIC) chipset for the XGIS instrument, named ORION, is presented.  Both ORION front-end (ORION-FE)
and ORION back-end (ORION-BE) structures with their interaction are described. Simulation results of a single-channel prototype, which is currently being experimentally characterized, are discussed.

\section{XGIS: GENERAL OVERVIEW}
\label{general_overview}

%cita Labanti

\begin{figure} [ht]
\centering
\begin{tabular}{c}
\includegraphics[height=7.5cm]{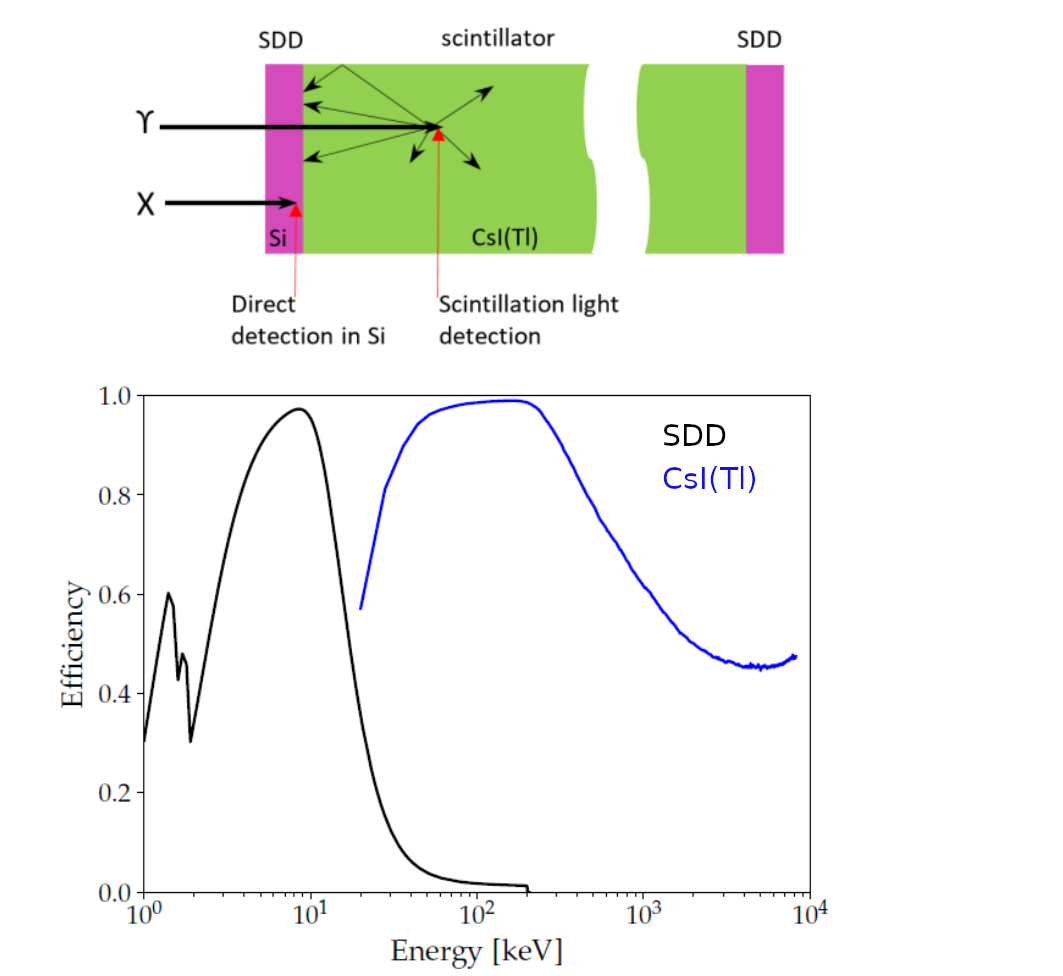}
\includegraphics[height=7.cm]{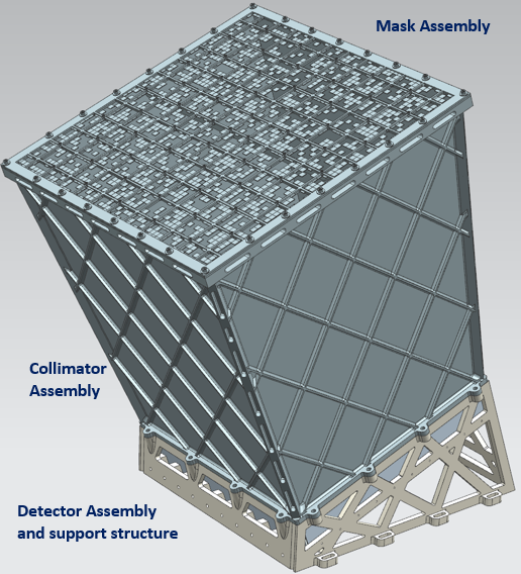}
\end{tabular}
\caption[example] 
{ \label{fig:xgis_princ_eff_mod}
\emph{Top left:} the working principle of a \emph{siswich} detection system. Two SDD are placed at both ends of a CsI(Tl). The low energy radiation (below $\sim$30 keV) interacts in the SDD. Photons with energy above a certain threshold cross the SDD and interact in the scintillator bar. The light output is collected at both SDD coupled with the CsI(Tl) bar. \emph{Bottom left:} efficency vs. photon energy for the direct detection in the SDD (black curve) and through light conversion in the CsI(Tl) (blue curve). \emph{Right:} design of one XGIS camera on board THESEUS.}
\end{figure}

The XGIS\cite{amati2020spieA} instrument is composed by two cameras which are tilted of $\pm$~20$^\circ$ in the two opposite directions with respect to the satellite axis\cite{mereghetti2020spie}~. To enable imaging capability for each XGIS Camera in the 2--150~keV energy range, a coded aperture system placed 63~cm above the detection plane will be used. In order to provide effective coding of the sky in the full energy range, the coded mask opaque elements are 1~mm thick. In its imaging energy range, the XGIS FoV fully overlaps that of the SXI. A collimator made of Al enclosing a W foil 0.25 mm thick limits the FoV of each XGIS Camera to 77$^\circ$~$\times$~77$^\circ$. Under these assumptions the overall FoV of the XGIS will be 117$^\circ$ $\times$ 77$^\circ$. 
Each Camera is composed by a matrix of 5 $\times$ 2 \emph{Super-Modules}, while each Super-Module is in turn composed of 10 \emph{Modules}. Finally, each Module 
consists of a matrix of 8~$\times$~8 pixels which are 
Thallium activated Cesium Iodide (CsI(Tl)) scintillator bars coupled to two Silicon Drift Detectors (SDD) cells, at top and at bottom side 
of each CsI(Tl) bar. The SDDs work both as direct detectors for low energy photons and
as a readout system for the scintillation photons. Indeed, low energy 
X-rays ($<$30~keV) interact directly in the SDD, while for energies above 30~keV the indirect photon detection occurs with the photon interaction in the CsI(Tl) crystals. 
The resulting scintillation light, which is in the optical energy range, is collected by SDDs on both side of the crystal.
Such configuration, allowing for a compact and efficient broadband single device detection unit, has been also called \emph{siswich} (from Silicon-sandwich) 
\cite{prete92, friese93, marisaldi05} in analogy with the phoswich detection system successfully used in several experiments 
(e.g. PDS/BeppoSAX\cite{frontera92}~, HEXTE/RXTE\cite{rothschild98}~, 
HE/HXMT\cite{liu2020}). The SDD size determines the spatial resolution of the detector. Furthermore, if the interaction occurs in the scintillator bar, by weighting the signals from the SDD 
at both ends (whose intensity is inversely proportional 
to the distance between the interaction point and the 
SDD), the coordinate of the scintillation point along 
the bar can be 
estimated. Figure~\ref{fig:xgis_princ_eff_mod} 
illustrates the working principle of the siswich detection system, together with the achievable detection efficiency.

In the XGIS configuration, each CsI(Tl) bar has 
dimensions 5~$\times$~5~$\times$~30~mm$^3$, the longer dimension along the the XGIS Camera axis. A custom electronic read-out will distinguish the two kind of signals, depending on the interaction 
that can occur directly on the topmost SDD (the one facing the sky) or on 
both SDDs, meaning that the radiation interacted in the CsI(Tl) bar.
In particular, electron-hole pair creation from X-ray 
interaction in Silicon generates a fast signal (about 
100~ns rise time). On the other hand, due to the fluorescent light component and crystal characteristics, the scintillation light collection is emitted and collected in a much longer timescale (typically 1--2 $\mu$s).

\begin{figure} [ht]
\centering
\begin{tabular}{c}
\includegraphics[height=7.5cm]{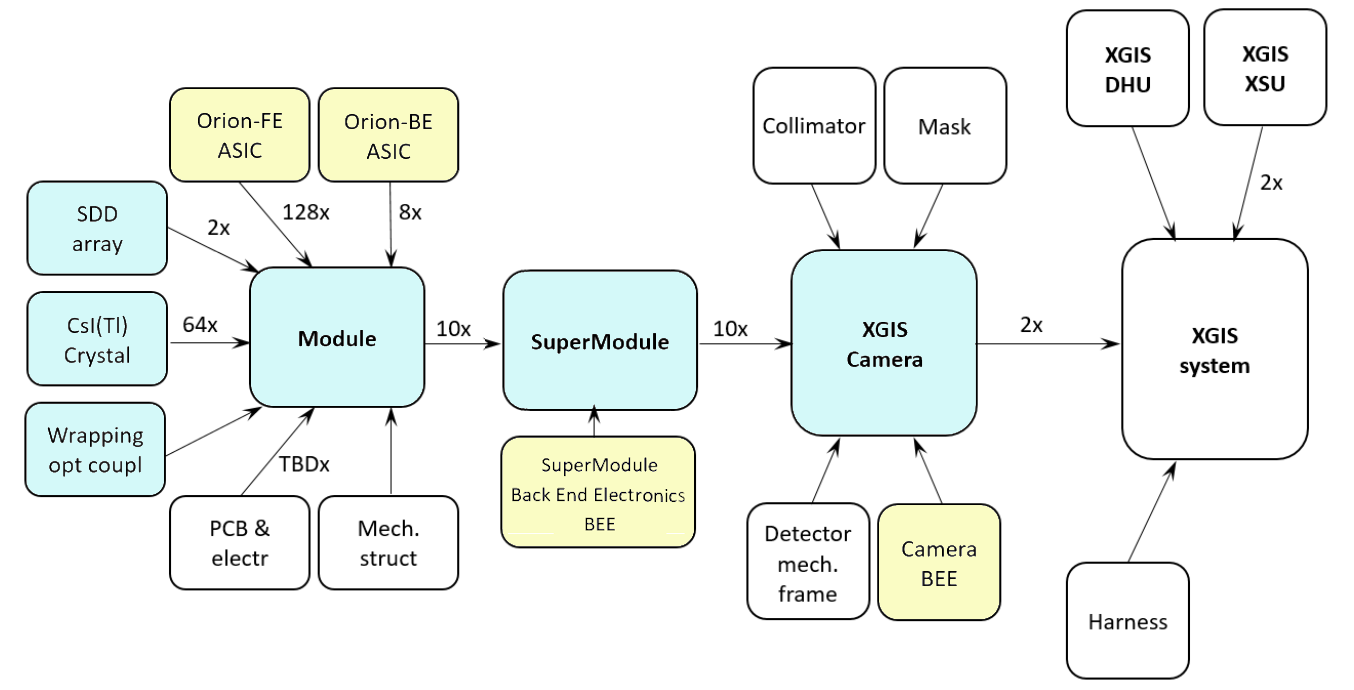}
\end{tabular}
\caption[example] 
{ \label{fig:xgis_structure} 
Block diagram showing the XGIS systems and subsystems. In the present paper we focus on the  modularity of the XGIS (cyan blocks) with increasing complexity from left to right and on the electronics (yellow blocks).}
\end{figure}

The segmented structure leading progressively from a  
single SDD+CsI(Tl)+SDD pixel to a camera (Figure~\ref{fig:xgis_princ_eff_mod}) and to 
the XGIS full system is also explained with the 
diagram shown in Figure~\ref{fig:xgis_structure} 
in which are also reported the sub-systems correlated to
each structure of increasing complexity. In particular, 
the highlighted blocks are those directly connected both
to the modularity of the instrument and to its 
associated electronics. These two topics are the 
subjects of this paper. 
In the following Sections we will 
describe the Module ORION ASIC, the Super-Module 
back-end electronics (BEE) and the Camera BEE, respectively.
Finally, a general description of the XGIS system will be reported with particular highlight to the the power distribution, to the Data Handling Unit (DHU) and to the total power consumption of each XGIS camera.

\section{XGIS MODULE silicon sensors: SDD}

Due to the the peculiarity of the detector architecture of the combined detection of X and Gamma radiation in the THESEUS XGIS instrument, a fully custom SDD matrix design is needed.
The SDD development activity is a natural progress of the developments carried out within the ReDSoX research program\footnote{\url{http://redsox.iasfbo.inaf.it}} financed by INFN and co-financed jointly by FBK for the part related to the SDD technology. 

In particular, the XGIS application provides the optical coupling of two SDD sensors to the opposite faces of a scintillator bar to detect the light generated by the interaction of gamma radiation within the crystal. In this configuration, the SDD sensor is also capable of detecting the X radiation impinging on the face not coupled to the crystal, facing the sky.
This solution is also suitable to the realization of a detector with imaging capability by packing together a certain number of scintillator bars read out by monolithic SDD sensor matrices. In the ReDSoX context, some SDD sensor prototypes dedicated to this application have been designed, implemented and characterized. Already in the early stages of prototype development, the size of the scintillator bar section was originally chosen as 5~mm $\times$ 5~mm. Consequently, SDDs have been designed with elements with this active area and the design parameters have been determined so as to optimize the performance for this specific dimension. The simulation and design of the devices is carried out by the INFN section of Trieste and by the University of Udine, while the realization takes place in the Micro Nano Facility of FBK. A first technological validation test of the sensors is performed in FBK at wafer level, while a more accurate characterization is carried out after the cut by TIFPA-TN in collaboration with INFN-TS. 

\begin{figure} [ht]
\centering
\begin{tabular}{c}
\includegraphics[height=5cm]{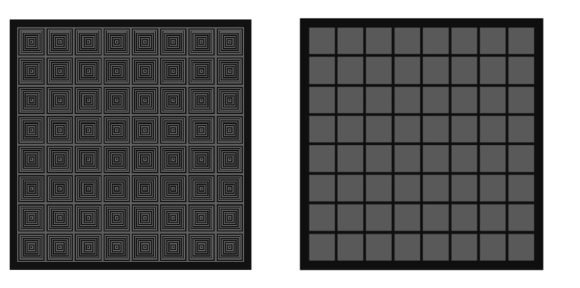}
\end{tabular}
\caption[example] 
{ \label{SDD} 
XGIS SDD array with 8$\times$8 individual SDDs. \emph{Left}: the $n$-side, with collecting anodes. \emph{Right}: the $p$-side that will be optically coupled to scintillators}
\end{figure}

In a XGIS module are employed 2 SDD-arrays, each one made by 8$\times$8 individual SDD elements as shown in Figure~\ref{SDD}, whose parameters are summarized in Table~\ref{SDDtable}. The detectors are made using a full double-side process: the p-side is the side in which the optical entrance window is made exploiting a shallow implant, while the n-side is the side in which the collecting anode electrode is made by an ohmic bulk contact, surrounded by the drift rings.
The basic geometrical parameters are: the pitch between elements, that has been fixed to 5 mm, and the peripheral guard-rings structure ($\sim$1.2 mm wide), whose represents an insensitive area to the incoming radiation, reducing the overall fill factor of the final assembly.

The $n$-side of the detector will be directly exposed to radiation. The sensitive area for X-rays detection, which in principle is the geometrical area 5$\times$5~mm$^2$ of each array element, is instead reduced due to a partial obstruction made by the assembly PCB. At the end, the final sensitive area will be defined by the definitive version of the assembly PCB. The $p$-side of the detector will be optically coupled to the scintillators. To guarantee an adequate optical separation between adjacent elements, a 0.5 mm wide Al track is deposited between the SDD elements, thus reducing the sensitive area for the scintillation light to 4.5$\times$4.5~mm$^2$.

\begin{table}[ht]
\caption{Main parameters of the SDD array}
\begin{center}
\begin{tabular}{@{} *2l @{}}
\toprule
%\emph{name} & \emph{foo} &&&  \\\midrule
%Models    & A  & B  & C  & D  \\ 
 Array size          & 42.4 $\times$ 42.4 mm$^2$\\       %$X$ & X1 & X2 & X3 & X4\\ 
 Si thickness        & 450 $\mu$m\\     %v& Y1 & Y2 & Y3 & Y4\\\bottomrule
 $\#$ of SDD            & 64\\
 Single SDD actrive aerea (n side)  & 5 $\times$ 5  mm$^2$\\
 Single SDD active area for scintillator (p side) &   4.5 $\times$ 4.5  mm$^2$\\
 Metal grid between single SDD (p side)  & 0.5~mm wide\\
 Typical polarization voltage (1 connection for the whole array)  &  --100 $\div$ --150 V\\
 Typical return voltage (1 connection for each SDD) &  --12 $\div$ --20~V\\
 Single SDD capacitance  & 50~fF (typical)\\
 Dark current (typical at T = 20$^\circ)$  & 50~pA (typical)\\
 Optical spectral response & 350 -- 1000~nm (typical)\\
 QE & $>$ 80\%\\
 \hline
\label{SDDtable} 
\end{tabular}
\end{center}
\end{table}

The depletion voltage of each SDD (VDEPL), is strongly linked to the doping concentration of the substrates and then for a large area detector the doping uniformity at wafer level is crucial. For devices manufactured in the same production run with the same substrate batch VDEPL uniformity level well fit the requirements for the production of the XGIS sensor for THESEUS. However for XGIS-Camera we foresee to select SDD with uniform VDEPL for the same Super-Module (assembly with 10 Modules mechanically and logically connected), regulating the VDEPL at Super-Module level via TLC.
A first production of such devices, to both check the process flow and the device layout, has been carried out in FBK during the first part of 2019 and concluded in June 2020. Such batch results in a multi-project wafer assembly. Preliminary measurements were performed in FBK to acquire the IV-total curve, with the aim to estimate the leakage current density for each SDD array, as a kind of production quality check. From such measurements, SDD arrays results in very low-level leakage current density, estimated to be $<$200~pA/cm$^2$ measured at +24$^\circ$C, then considered very promising for further investigations and characterization of the single SDD cells.

\begin{comment}
\begin{figure} [ht]
\centering
\begin{tabular}{c}
\includegraphics[height=5cm]{images/SDDtable.png}
\end{tabular}
\caption[example] 
{ \label{SDDtable} 
Main parameters of the SDD array}
\end{figure}
\end{comment}

\section{XGIS MODULE READOUT ASIC: ORION}
\label{module_asic_orion}

\begin{figure} [ht]
\centering
\begin{tabular}{c}
\includegraphics[height=6cm]{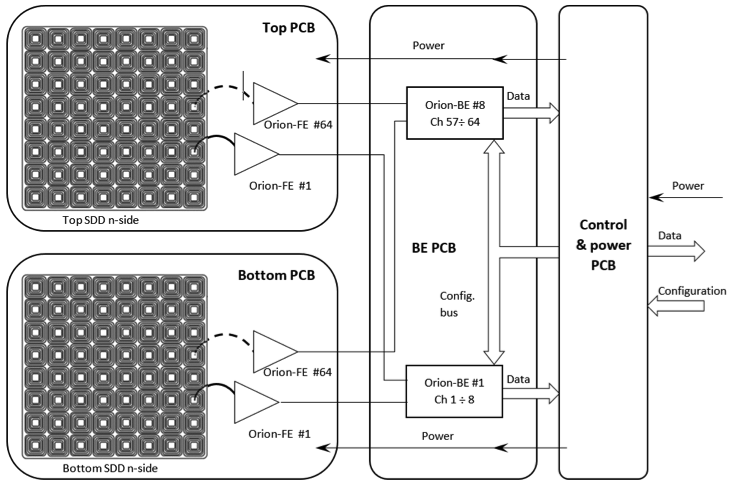}
\end{tabular}
\caption[example] 
{ \label{fig:arcORION} 
Structure of the electronics for collecting and processing the SDD signals based on the ORION-FE and ORION-BE ASICs.}
   \end{figure}

The electrical signals from the SDD anodes are collected and processed by the ORION chipset, a constellation of ASICs composed by the ORION-FEs, for the charge readout and the initial signal shaping placed in close proximity of the SDDs, and the ORION-BEs for the complete signal processing and digitization. The simplified schematic of the ASIC readout architecture is shown in Figure~\ref{fig:arcORION}. A similar structure is adopted in the HERMES (\emph{High Energy Rapid Modular Ensemble of Satellites}) nano-satellites mission\cite{burderi18, fuschino19} in which the ASICs of the LYRA family (LYRA-FE and LYRA-BE) are employed.

The ORION chipset is responsible for both the analog readout of the SDD charge and for the digitization and data communication of the event information to the module electronics. For each detected event, the ORION ASIC is able to provide the energy of the event, the type of the event (i.e. X or $\gamma$) the position of the event (i.e. pixel coordinates) and the timing of the event with respect to an external clock. 
The ORION chipset is composed by a total of 
12800 analog ORION-FE that send a pre-shaped signal to 800 mixed-signal 
back-end multi-channel chips (8-channels ORION-BE) for dedicated signal processing and digitization. 

\begin{figure} [ht]
\centering
\begin{tabular}{c}
\includegraphics[height=7cm]{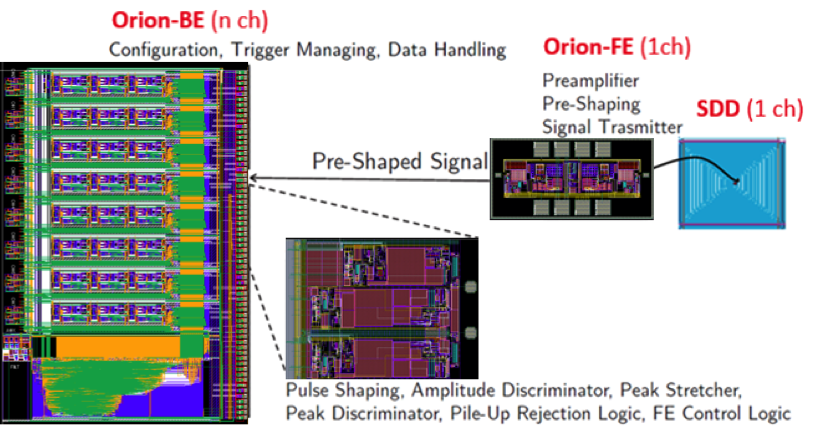}
\end{tabular}
\caption[example] 
{ \label{fig:asic_architecture1} 
Structure of the electronics for collecting and processing the SDD signals based on the ORION-FE and ORION-BE ASICs.}
\end{figure}

Due to the peculiar architecture of the ASIC, each 
single channel ORION-FE is physically placed 
close to the SDD anode, in order to keep the stray 
capacitance 
at the preamplifier input as low as possible. 
The ORION-FE single channel collects the SDD generated 
charge and performs pre-amplification. The ORION-BE, that is 
physically placed a few cm away on the bottom of the 
module, receives the signals from 
8$\times$2 ORION-FEs, processes them, handles the time marking of each single event and interfaces with 
the module electronics (Figure~\ref{fig:asic_architecture1}).

\begin{figure} [ht]
\centering
\begin{tabular}{c}
\includegraphics[height=7cm]{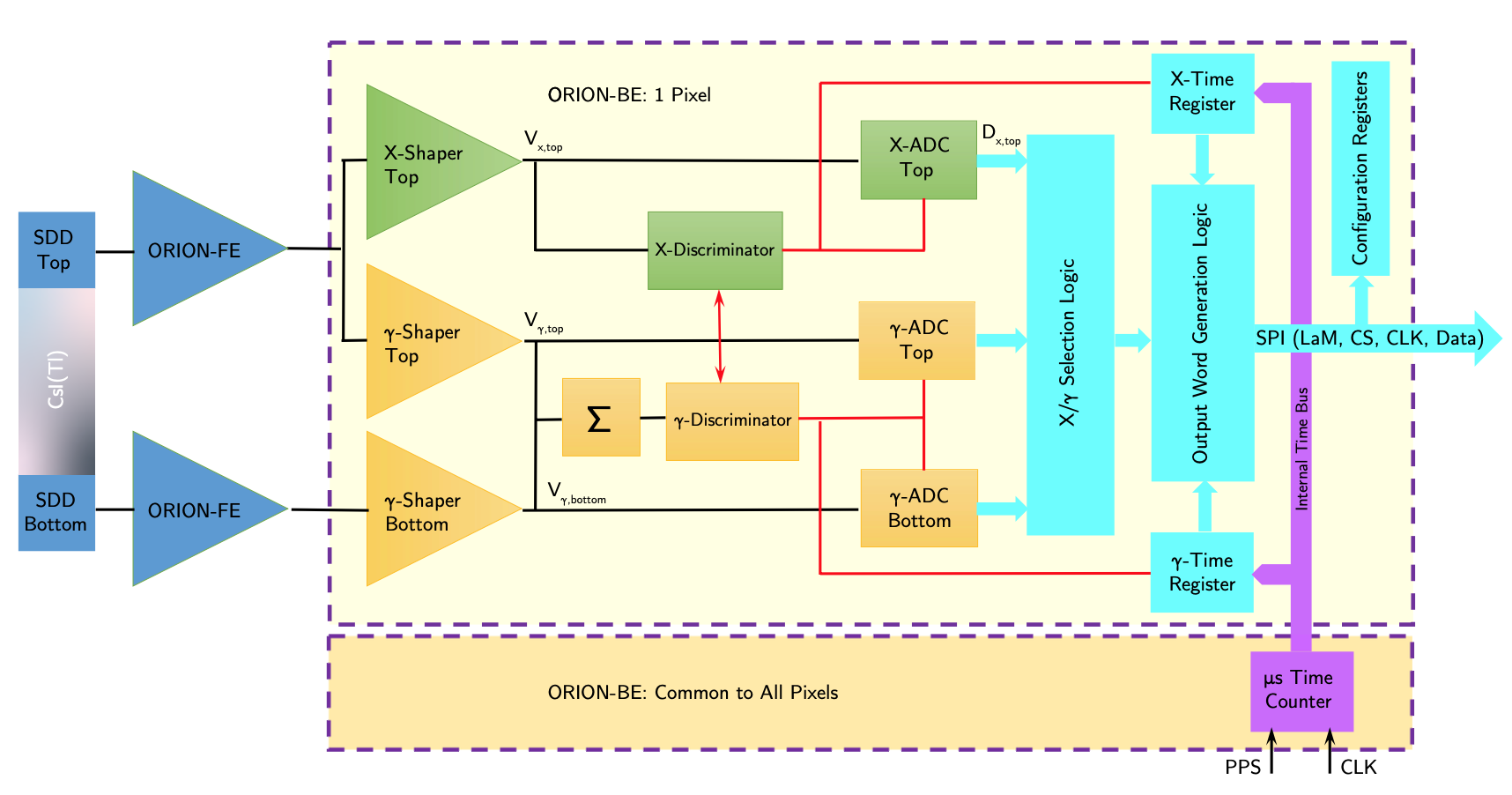}
\end{tabular}
\caption[example] 
{ \label{fig:asic_architecture2} 
The ASIC architecture related to a pixel.}
\end{figure}

\begin{figure} [ht]
\centering
\includegraphics[height=3.5cm]{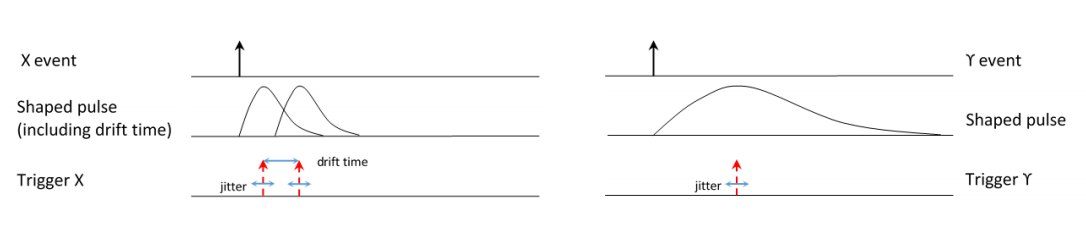}
\caption[example] 
{ \label{fig:ORIONtiming} 
The time marking of the X and $\gamma$-events. $\it Left$: an X event is detected only in the top SDD but the time tag is affected by both the trigger jitter combined with the uncertainty due to the charge drift time due to the unknown interaction position within the SDD cell. $Right$: the time tag of a $\gamma$-event depends on the trigger jitter combined with a longer signal collection due to the fluorescence light component.}
\end{figure}

Figure~\ref{fig:asic_architecture2}  shows the general ASIC architecture for a single pixel, which, as discussed previously, is composed by two SDD optically coupled with a CsI(Tl) bar at both ends. The rationale for this architecture is 
the following:

\begin{itemize}

\item If an X-ray (2--30~keV) is detected only in the top SDD (hereafter called \emph{X-mode event}):
\begin{enumerate}
    \item The preamplifier output signal is fast (hundreds of ns).
    \item As the X-ray detection is (almost) point-like in the SDD, the arriving time of the signal at the SDD anodes is delayed up to 1~$\mu$s with respect to the event occurrence, due to the drift time of the 
    electron cloud within the SDD. The time marking of the event is then affected by the uncertainly due both to the jitter of the trigger and to the unknown position of the interaction in the 
    SDD (Figure~\ref{fig:ORIONtiming}).
    \item The best signal/noise ratio is achieved with a short shaping 
    time (1~$\mu$s typical).
    \item The discrimination between X-ray and $\gamma$-ray is done only 
    using the top SDD signal (see below).
\end{enumerate}
\item If a $\gamma$-ray ($>$20 keV) is detected simultaneously in top 
and bottom SDDs (hereafter called \emph{$\gamma$} or \emph{S-mode event}):
\begin{enumerate}
    \item The preamplifier output signals is slow (of the order of few 
    $\mu$s), in agreement with the typical scintillation decay time.
    \item The scintillation light is spread across the whole SDD, therefore 
    both scintillation time and maximum drift time of the charge should 
    be taken into account to avoid ballistic deficit.
    \item The best signal/noise ratio is achieved with shaping time 
    of the order of 3~$\mu$s (typical).
    \item The amplitude discrimination is achieved operating on the sum 
    of top and bottom SDD signals.
\end{enumerate}
\item Mixed X-ray/$\gamma$-ray events in time coincidence can occur with quite low probability (few \% even at the highest photon energies).
\end{itemize}

%\begin{comment}
%\newline
%The signals from the SDDs are so processed by:\newline
%•	ORION-FE ASIC: collect the SDD charge signal, operating a %charge-voltage conversion and a first filtering step;\newline
%•	ORION-BE ASIC: receive the signals from 8×2 ORION-FE, process %them, handle the time marking of single events and interface with the module electronics;\newline
%\end{comment}

%The structure of the electronics for collecting and processing 
%the single pixel signals based on the ORION-FE and ORION-BE 
%ASICs is presented in Figure~\ref{ORIONChips}.

%\begin{figure} [ht]
%\begin{center}
%\includegraphics[height=6cm]{images/ORIONChips.png}
%\caption[example] 
%{ \label{ORIONChips} 
%Structure of the electronics for collecting and processing the single pixel  signals based on the ORION-FE and the ORION-BE (8-channel version) ASICs}
%\end{center}
%\end{figure} 

\subsection{ORION-FE}
\label{orion_fe}

The ORION-FE is a fully-analog ASIC conceived to provide the first amplification of the charge signal coming from the SDD, keeping a low area occupation ($\sim$0.3 mm$^2$) in order to allow a close connection to each anode, without compromising the detection area. The ORION-FE is mainly divided in three stages: a Charge Sensitive Amplifier (CSA) with a dynamic range of 32~fC, followed by a pole-zero compensation stage and a current conveyor, which additionally introduces a first CR shaping with 1 $\mu$s time constant.  To avoid spurious injection the CSA is operated in continuous reset mode, and the output of the ORION-FE is delivered as a current signal generated by the current conveyor itself . Each ORION-FE has a power consumption of 290 $\mu$W at T = $-$20 $^\circ$C, which also includes the contribution from the internal biasing module and of the Electrostatic Discharge (ESD) protection module.

\subsection{ORION-BE AND LOGIC}
\label{orion_be}

The signals generated by the top and bottom ORION-FE of each pixel are transmitted as current signals on $\sim$4--5~cm long lines and readout by the ORION-BE using two low-input impedance current receivers. A signal coming from the bottom SDD is uniquely identifying a $\gamma$-ray event, 
while the signal coming from the top SDD cannot be a priori identified 
as an X or $\gamma$-ray event, and the output of the top current receiver is sent to both X-/$\gamma$ processing channels, which operate with peaking times of 1 $\mu$s and 3~$\mu$s, respectively. The stretched output voltage is then digitized by a second-order 12 bit ENOB incremental A/D converter. After the conversion, the pixel logic is able to discriminate the type of detected event, with the respective timestamp, and eventually forward the information to the on-board module electronics. Each read-out BE channel is thus made up by three parallel paths, each one including also an independent ADC. Even if the BE discriminators deliver X and $\gamma$ related triggers on a given event, due to the nature of the instrument, this information is fundamental but not sufficient to determine the type of event. For this reason the ADC outputs as well as the triggers from the BE are processed by a dedicated logic which has the double aim of extracting the type of event, assigning the timestamp and creating the digital output frame. This frame will include information on the event time, the type of event, its address, the discriminators triggered during the event and, in probe mode, further information useful for detailed characterization of the ASIC, such as, for instance, the signals on the internal ADC buses. In probe mode through a multiplexer and a buffer also all the analog outputs of the BE channels and their triggers are available for external measurement. Finally, in the multi-channel ASIC several channels may trigger together for a common event. For this reason, the embedded logic waits a programmable time, named RTP (Rise-Time Protection), before sampling all the BE signals (analog and triggers), starting the data conversion and finally building the output frame. When the frame is ready, a Look at Me (LaM) signal is given. The ORION ASIC embeds 1-kbit memory bank for SDD read-out analog channel configuration, for internal logic settings and for temporary output frame storage. The configuration and the output frames can be loaded/read-out either through a shift register or an SPI interface, based on a 4-wire bus plus reset. The ORION-BE has a power consumption of 980~$\mu$W per pixel (including both X/$\gamma$ channels) at T = $-$20 $^\circ$C.

\subsection{ORION RESULTS}
\label{simulation_results}

%\textcolor{red}{Se eventualmente Mele, Malcovati, Grassi, Bertuccio vogliono integrare.}

\begin{figure} [ht]
\centering
\includegraphics[height=6cm]{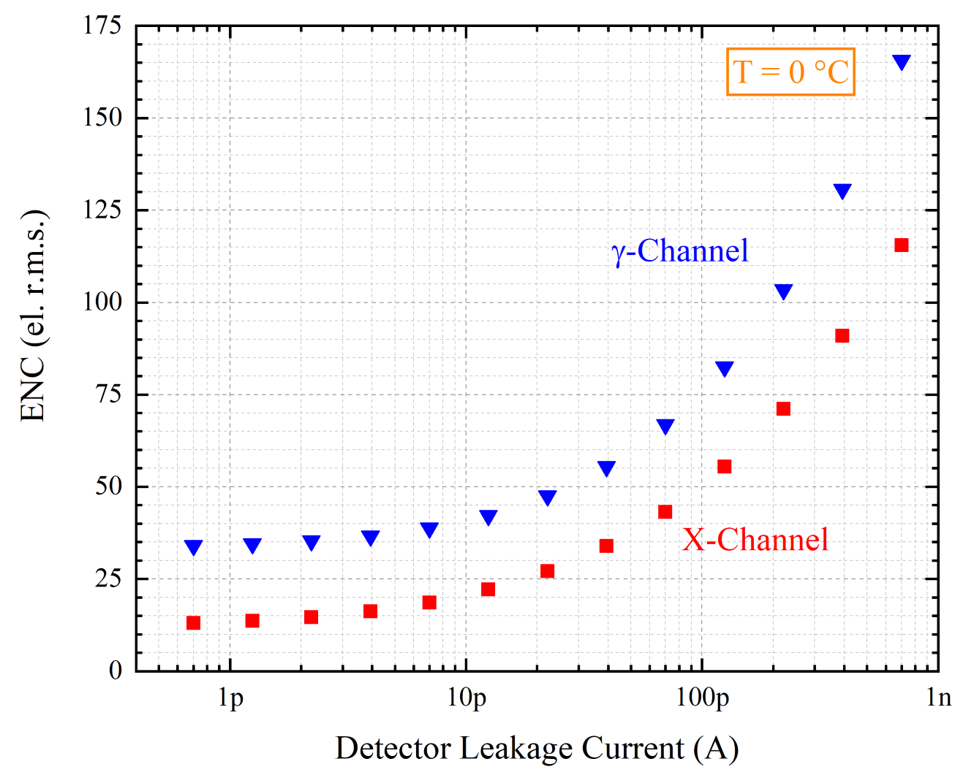}
\caption[example] 
{ \label{fig:ORIONenc} 
Simulated noise for different input leakage currents, corresponding to the expected Beginning-Of-Life (0.7~pA) and End-Of-Life (700~pA) performance of the detector.}
\end{figure}

\begin{figure}[th]
\centering
\includegraphics[height=4.5cm]{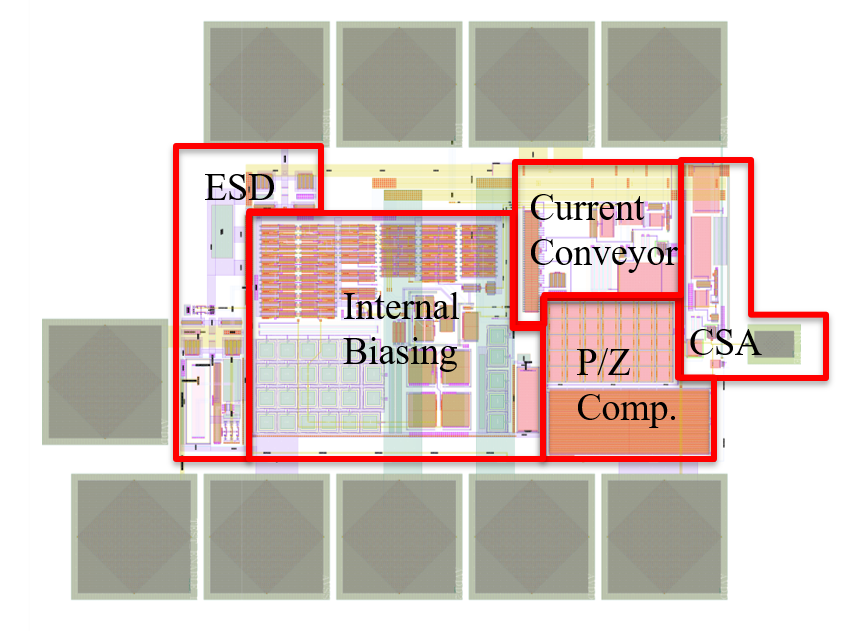}
\includegraphics[height=5cm]{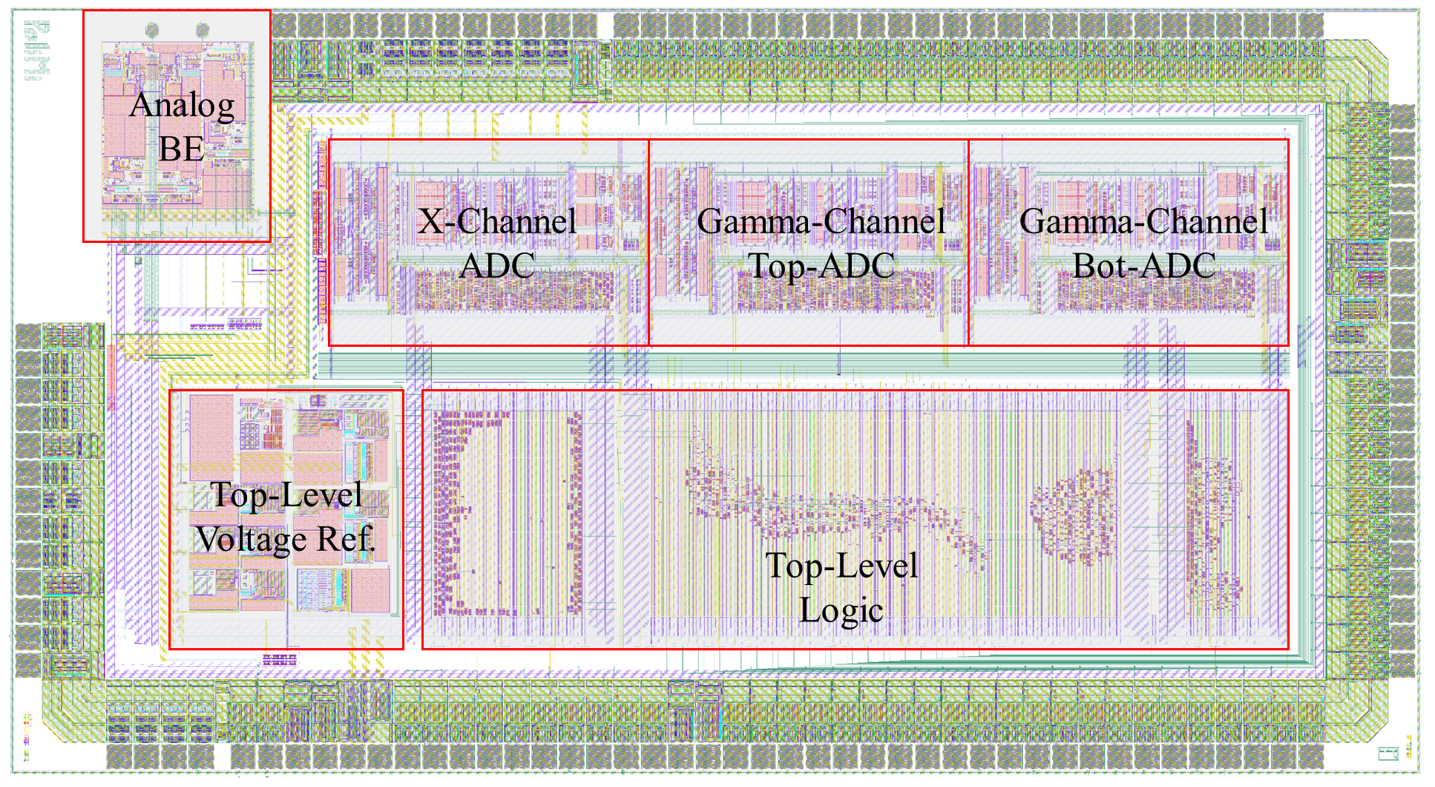}
\caption{\label{fig:ORIONlayout} A detailed layout view of the ORION-FE (left) to be placed in close proximity of the SDDs and of the single channel prototype of the ORION-BE (right). In the top-left corner of the ORION-BE, the analog BE module, with the two small-pad inputs, is visible. The analog-BE as well as the ADCs are realized on a modular layout structure, for ease of implementation in multi-channel architectures.}

\end{figure}

\begin{figure} [ht]
\centering
\includegraphics[height=6cm]{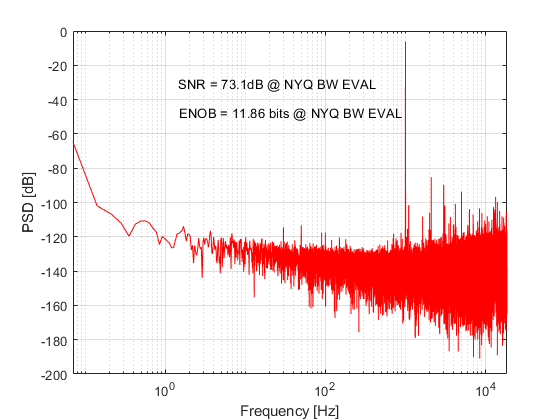}
\caption[example] 
{ \label{fig:ADC} 
ORION ADC elementary module output spectrum applying a 1~kHz sinusoidal signal at full scale}
\end{figure}

The ORION chipset has been simulated to have an Equivalent Noise Charge (ENC) (with a detector capacitance of 50 fF and a leakage current of 0.7 pA) of about 12.5 e$^-$ r.m.s. at 1 $\mu$s of peaking time on the X channel and 32.9 e$^-$ r.m.s. at 3 $\mu$s on the $\gamma$ channel. Despite a very wide input dynamic range on the $\gamma$ channel (from 400 e$^-$ to 180\,000 e$^-$) the expected linearity error is below $\pm$1.2\%, and below $\pm$0.1\% on the X channel. Due to the expected degradation of the SDD leakage current during the whole mission lifetime, the ORION chipset must be capable of handling high levels of input currents, up to 700 pA, which represents the worst case scenario for the detector leakage current in the End-Of-Life (EOL) condition, as shown in Figure~\ref{fig:ORIONenc}. The ORION chipset, whose architecture is shown in Figure~\ref{fig:asic_architecture1}, is currently under production in its 4-channel version, and the experimental characterization on one front-to-back channel is expected by the end of 2020. In fact, a single channel version of the ORION ASIC, the layout of which is shown in Figure \ref{fig:ORIONlayout}, is under experimental characterization, while a stand-alone analog-to-digital (A/D) converter module has been already produced and characterized. Figure~\ref{fig:ADC} shows the measured accuracy in terms of Effective Number of Bits (ENOB) of the elementary A/D converter developed \emph{ad hoc} for the ORION ASIC, featuring 11.9 bits including both noise and distortion effects. 

\section{SUPER-MODULE AND CAMERA BACKEND ELECTRONICS}
\label{supermodule-bee}

\begin{figure} [ht]
\begin{center}
\begin{tabular}{c}
\includegraphics[height=5cm]{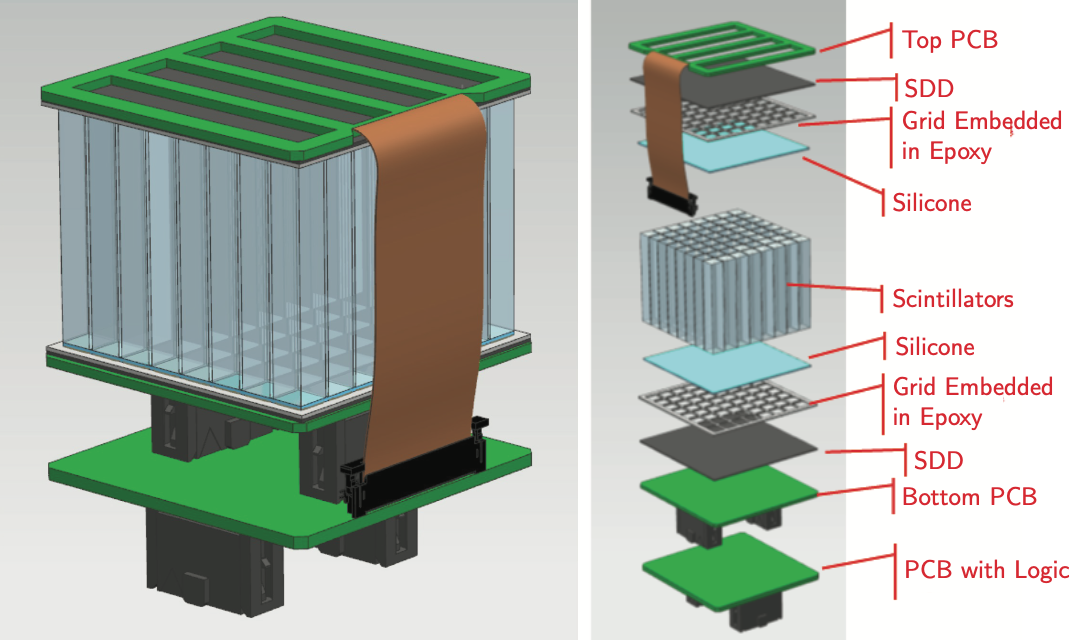}
\includegraphics[height=5cm]{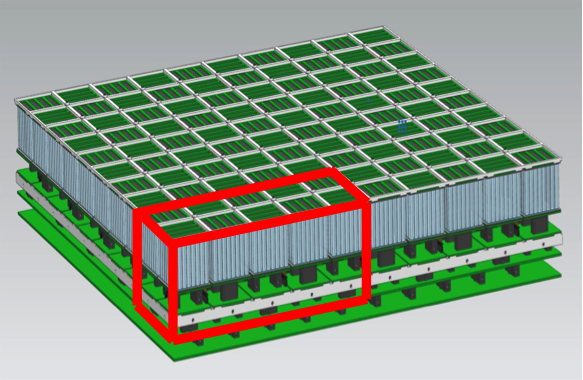}
\end{tabular}
\end{center}
\caption[example] 
{ \label{SuperModule} 
$Left$: A Module of the XGIS in which are visible the different parts that compose the detector pixels and the PCBs. $Right$: 
an XGIS Super-Module (highlighted in red)  which is made with 10 Modules. One entire XGIS camera, which is also visible in figure, is made with 10 Super-Modules.}
   \end{figure}

A XGIS Super-Module is a logical and hardware subset of the detector assembly and consists of 10 Modules joined with a Super-Module Back End Electronics board (SM-BEE). 
The segmentation of the XGIS camera in Super-Modules allows 
to mitigate the effects of a potential system failure that, 
in case of occurrence at Super-Module level, would consist 
in a reduction of only 10$\%$ of the sensitive area. 
The modularity at the ASIC 
level ensures that in case of failure of an ASIC only 
8 pixels are unusable. By exploiting the same 
concept, for each detected event, only 8 pixels get 
``frozen'' leading to a dead time interval (since the corresponding ASIC is busy). On the contrary, it allows to perform parallel acquisition of contemporary events falling in non-adjacent pixels. The segmentation in Super-Modules 
also ensure a remarkable reduction of the connections between ORION-BE ASIC and FPGA 
(5 lines), of the total FPGAs used in a XGIS Camera (10 FPGAs) 
and of components (ADCs+Logic embedded into the ASICs).
For each XGIS-Camera is also planned to install selected SDD with 
uniform depletion voltage (VDEPL) in the same Super-Module, with the possibility of regulating the VDEPL at Super-Module level via telecommands. The possibility, for large area detectors, of a fine tuning of VDEPL is crucial.
In Figure~\ref{SuperModule} are shown a general view as well as 
an exploded view of a single Module together with the 
full detection plane in which is highlighted 
its partition in 10 Super-Modules.

\begin{figure} [ht]
\begin{center}
\begin{tabular}{c}
\includegraphics[height=6cm]{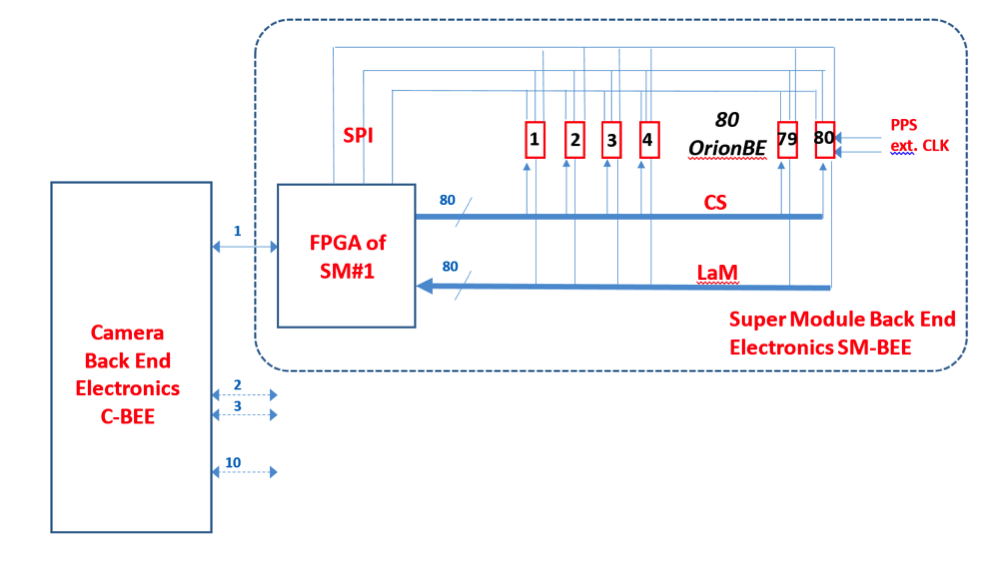}
\end{tabular}
\end{center}
\caption[example] 
{ \label{SMlogic} 
Logic concept of the Super-Module Electronics. One FPGA manages 80 ORION-BE and 10 SM-BEE are connected to the Backend Electronics of the Camera (C-BEE).}
   \end{figure}

The SM-BEE provides the power supply to 10 Modules, commands the Modules, collects the data (events/HKs) from the Modules and interfaces all the functions (commands, HKs, Alarms, Data) with the XGIS Camera Back End Electronics (C-BEE).
The architecture of the SM-BEE will be built around one FPGA 
that will control 80 Orion-BE ASICs with a logic
as depicted in Figure~\ref{SMlogic} whose main 
characteristics are: 

\begin{itemize}
\item for the I/F with each single ORION-BE ASIC there are only 5 logic lines between the FPGA (where 3 of them are in common);
\item the FPGA of the SM-BEE manages 160+3 = 163 logic lines in total (I/O pins);
\item the I/F with the C-BEE will be buffered with a memory of proper size;
\end{itemize}

\begin{figure} [ht!]
\begin{center}
\begin{tabular}{c}
\includegraphics[height=4cm]{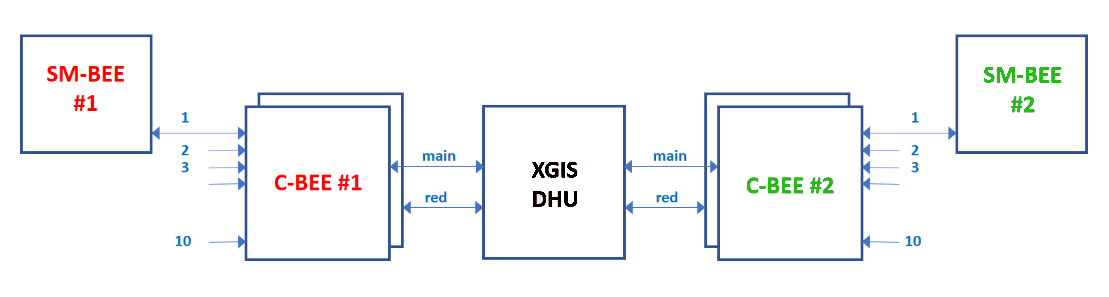}
\end{tabular}
\end{center}
\caption[example] 
{ \label{XGISconn} 
Logical connection between Super-Modules and XGIS-DHU through C-BEEs}
   \end{figure}

The XGIS Camera is managed through a Camera Back End Electronics board (C-BEE) which is organised in two logical sections, in cold redundancy, each connected to the DHU  main and redundant section respectively, and simultaneously connected with the 10 SuperModules. The main functions of the C-BEE are of providing power supply distribution and management of all Super-Modules and of power supply interface with the XSU.
Other functions of the C-BEE are: Data I/F with all Super-Modules, Data buffering and I/F with XGIS-DHU, telecommands I/F with XGIS-DHU, telecommands implementation and verification, HK management and transmission to XGIS-DHU and Alert management. The logical connection between SuperModules and XGIS-DHU through C-BEE is shown in Figure~\ref{XGISconn}.

\section{XGIS Electrical Architecture}
\label{electrical_architecture}

\begin{figure} [ht]
\begin{center}
\includegraphics[height=9cm]{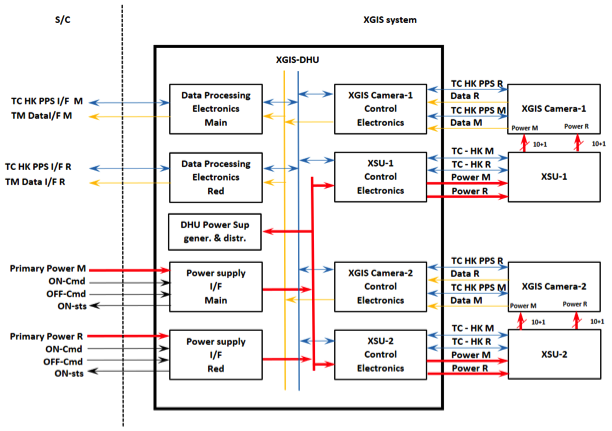}
\end{center}
\caption[example] 
{ \label{fig:xgis_module} 
Functional block diagram of the XGIS system, consisting of 2 identical cameras. Each camera is made of 100 Modules organized in 10 Super-Modules), 2 power distribution boxes (XSU) and a Data Handling Unit (DHU) (in cold redundancy).}
   \end{figure}

Beside the 2 XGIS-Cameras, two 2 XGIS Supply Units (XSUs) and a DHU compose the XGIS instrument. The XSUs will be contained in two boxes 
each one supplying one XGIS-Camera. The electrical architecture of the XGIS is shown schematically in Figure~\ref{fig:xgis_module} in which functional blocks of XGIS instrument and their interfaces with the 
THESEUS spacecraft platform are shown. Central element in the XGIS operation and data acquisition chain is the DHU that serves as telecommand/telemetry and power interface between the Spacecraft 
Service Module (SVM) and the XGIS-Cameras.
The on-board burst trigger capability is implemented as a part of 
the DHU, which is directly interfaced with the on-board data handling (OBDH) system. The electrical interface is assumed to be SpaceWire. 
Each C-BEE is connected to the DHU for control, data, and health monitoring. The bus power is routed through the XSU power 
distribution unit providing ON/OFF switching and protection 
capability, with an overall power consumption of about 210~W 
(included margins) for each camera.

\section{Summary and prospects}

In this paper we have shown the main features of the XGIS on board THESEUS. In particular we have described the ASIC ORION-FE and ORION-BE which has been developed specifically for the THESEUS mission concept. The concept of having the electronic chain split in two ASICs as in the LYRA FE and BE case has been tested
with success in the context of the HERMES nano-satellite mission. First prototypes of the ORION-FE and ORION-BE ASICs have been assembled and are currently in testing and characterization. 

\acknowledgments % equivalent to \section*{ACKNOWLEDGMENTS}       
The authors wish to thank the European Space Agency for its support through the M5/NPMC Programme and the Italian Space Agency and the National Institute of Astrophysics for their support through the ASI-INAF Agreement n. 2018-29-HH.0., the OHB Italia/INAF-OASBo Agreement n.2331/2020/01.

\bibliography{report} % bibliography data in report.bib
\bibliographystyle{spiebib} % makes bibtex use spiebib.bst

\end{document}